\documentclass[fleqn,10pt,twoside]{article}

\usepackage[english] {babel}

\usepackage[T1]{fontenc}
\usepackage[utf8]{inputenc}

\usepackage{graphicx}
\usepackage{graphics}
\usepackage{epsfig}
\usepackage{floatflt}

\usepackage {caption3}
\usepackage[labelsep=period]{caption}[2007/11/04 v.3.1e]

\usepackage{amssymb}
\usepackage{amsmath}
\usepackage{amsfonts}
\usepackage{euscript}

\usepackage {latexsym}

\usepackage {caption3} 
\usepackage[labelsep=period]{caption}[2007/11/04 v.3.1e]

\usepackage{cuted}  

\def\noi{\noindent}

\renewcommand{\thesubsubsection}%
        {\arabic{section}.\arabic{subsection}.\arabic{subsubsection}.}

\newcommand{\heads}[2]{\markboth{\protect\small\it #1}{\protect\small\it #2}}

\newcommand{\Arthead}[5]{ \setcounter{page}{#4}\thispagestyle{empty}\noi
    \unitlength=1pt \begin{picture}(500,40)

        \put(0,58){\shortstack[l]{\small\it ISSN 0202-2893, Gravitation \& Cosmology,
                          #1, Vol.#2, No. #3, pp. #4--#5    
\footnotesize\copyright \ Pleiades Publishing Ltd., 2018.} }
    \end{picture}
	 }     		
\def\prepno#1#2
    {\thispagestyle{empty}
    \noi    \unitlength=1mm
    	\begin{picture}(178,10)
            \put(177,15){\llap{\bf #1}}
            \put(177,10){\llap{\small\rm #2}}
        \end{picture}
          }             

\newcommand{\Title}[1]{\noi {\uppercase{\Large #1}}     }
\newcommand{\Author}[2]{\noi{\large\bf #1}\\[2ex]\noindent{\it #2}   }

\newcommand{\Abstract}[1]{\vskip 2mm \begin{center}
        \parbox{16.4cm}{\small\noi #1} \end{center}\medskip}

\newcommand{\foom}[1]{\protect\footnotemark[#1]}

		\newcommand{\email}[2]{\footnotetext[#1]{E-mail: #2}
		\addtocounter{footnote}{1}}

\heads{A.M.Baranov}
      {OPEN UNIVERSE MODELS ...} %
 
\begin{document}

\twocolumn 
[ 
\Arthead{2018}{24}{4}{344}{349.}
\vspace{-10mm}


\begin{center}
\Title{\bf  Open Universe Models as Analogs \\
\vspace{0.2cm}
of Multidimensional Electric Capacitors}
\end{center}

\begin{center}
\vspace{.5cm}
   \Author{A.M.Baranov\foom1}   
{\it Astafyev Krasnoyarsk State Pedagogical University, \\
 ul. Ady Lebedevoy 89, Krasnoyarsk, 60049, Russia ; \\
Reshetnev Siberian State University of Science and Technologies (SibSAU), \\
prosp. gazety ''Krasnoyarskiy rabochiy'', Krasnoyarsk, 60037, Russia}

{ \footnotesize{Received March 27, 2018}}
\end{center}

\Abstract
{{\bf Abstract}--A new approach to obtaining open Universes models as exact solutions of gravitational equations is considered. The proposed method is based on an analogy between electrostatics of conductors and open cosmological models which have a conformally-flat 4-metric in the Fock form. 
 These cosmological models are solutions of the Einstein equations with the energy-momentum tensor of a Pascal perfect fluid. As a result, it is shown that the capacity of the multidimensional "ball" electric capacitor in multidimensional Euclidean auxiliary space is connected with harmonic functions by which the potentials of multidimensional capacitors are described. In turn, the conformal factor of the metric describing an open universe model in the Fock representation is a power function of this potential for each dimensionality of the Euclidean space. In particular, for 3D auxiliary Euclidean space in 4D space-time, there is the Friedman solution for an open universe filled with incoherent dust (with flat metric at spatial infinity). Further for 4D Euclidean space, we have an analogue of the open cosmological model with the equation of state of an ultrarelativistic gas. Other cases are listed in a table. A table of the matter states which generalizes this approach to multidimensional space-times is also constructed. Thus, the possibility of replacing the modelling problem in cosmology with the equivalent electrostatic problem for finding the capacity of capacitors is shown under specified boundary conditions. Exceptions from the general approach are also presented in the article.
}

{\bf DOI:} \tt\small{ 10.1134/S0202289318040059}
\vspace{1.0cm}

] 

\vspace{-100.0cm}

\email 1 {alex\_m\_bar@mail.ru}

\begin{center}
\section{\large{\tt INTRODUCTION}}
\end{center}

\indent
The approach to deriving open universe models with various physical states is based, as a rule, on the solution of the gravitational equations with specific linear equations of state, as it is shown in 
[1--3]. Another approach can be seen in [4--6]. This study is an attempt to consider such cosmological models fron another viewpoint, using their analogies with electrostatics of conductors. The conductors create a multidimensional system which can be named a multidimensional electric capacitor, or $k$-capacitor. To begin with, a metric conformal to the Minkowski metric in 4D space-time is 
\begin{equation}
d{{s}^{2}}\ = e^{2\sigma }{{\eta }_{\mu \nu }}\,dx{{\,}^{\mu \,}}dx{{\,}^{\nu }},
\end{equation}
where $e^{2\sigma }\,$ is a conformal factor which depends on a variable $S$ only,
$\sigma =\sigma (S)$; ${{S}^{2}}={{t}^{2}}-{{r}^{2}}=\ {{\eta }_{\mu \nu }}\,{{x}^{\mu }}\,{{x}^{\nu }}\,$; $\ {{\eta }_{\mu \nu }}\,=\ diag(1,-1,-1,-1)\,;\ \mu ,\,\nu \ = 0,\,1,\,2,\,3\,$; $t$ and $r$ are the time and radial variables.

The metric (1) is written in Fock's metric form [7] for open cosmological models and describes space-times with negative curvature. The Einstein equations are written without cosmological term:
\begin{equation}
R_{\mu \nu }-{{g}_{\mu \nu }}R=-\varkappa T_{\mu \nu },
\end{equation}
and with the matter source of the gravitational field the energy-momentum tensor (EMT) in an approximation of Pascal perfect fluid
\begin{equation}
{{T}_{\mu \nu }}=\varepsilon \,{{u}_{\mu }}{{u}_{\nu }}+p\,{{b}_{\mu \nu }}
\end{equation}
with a linear equation of state
\begin{equation}
p=\beta \cdot \varepsilon ,
\end{equation}
where ${{R}_{\mu \nu }}$ is the Ricci tensor; $R$ is the scalar curvature; $\varkappa $ is the Einstein gravitational constant; $\varepsilon $ is the energy density; $p$ is the pressure; $\beta =const $ and for energy dominance $|\beta | \leq 1$; ${{u}_{\mu }}=(\exp (\sigma ))\cdot {{b}_{\mu }}$ is a 4-velocity; ${{b}_{\mu }}\equiv S{{,}_{\mu }}$; ${{b}_{\mu \nu }}=({{u}_{\mu }}\,{{u}_{\nu }}\ -{{g}_{\mu \nu }})$ is a 4-projector or the 3-metric tensor orthogonal to $\,{{u}^{\mu }},$ $\;{{b}_{\mu \nu }}{{u}^{\mu }}=0;\;$ the velocity of light and the Newton gravitational constant are taken to be equal to unity.

As the parameter $\beta $ runs through a number of values satisfying the energy dominance principle $|\beta | \leq 1,$ we have some interpretations of matter equations of state (see, for example, 
[3] and Table 1). Further, the Einstein equations can be split by projecting the physical quantities onto the timelike direction and 3D surface which is orthogonal to this direction. It is the method of 
(3+1) splitting in general relativity [8,9] for finding physical 

\newpage

\begin{strip}

\vspace{-9mm}

{\bf Table 1.} States of matter in multidimensional space-time with asymptotically flat  metrics for open cosmological models

\begin{center}
\begin{tabular} {l|c|c|c|r}     \hline
 $k$ & $k-1$ & {$\beta=p/{\varepsilon}$}& {State of the matter}& {Conformal factor} $\exp(2\sigma(S)$)                             \\ \hline
 0   & -1    & -1     & physical vacuum & ${{(B+A\cdot {{S}^{2}})}^{-2}}$ \\ 
 1   &  0    & -2/3   & domain walls    & ${{(B+A\cdot S)}^{-4}}$         \\ 
 2   &  1    & -1/3   & relativistic strings & $A\cdot {{S}^{-2B}}$        \\
 3   &  2    &   0    & incoherent dust  & ${{(B+A/S)}^{4}}$             \\ \ 
 4   &  3    &  +1/3  & ultra-relativistic gas & ${{(B+A/{{S}^{2}})}^{2}}$  \\ 
 5   &  4   &  +2/3  & non relativistic degenerate gas & ${(B+A/{{S}^{3}})}^{4/3}$ 
\\ 
 6   &  5    &  +1  & superstiff & $(B+A/{{S}^{4}})$          \\ \hline
\end{tabular}
\end{center}
\vspace{-5mm}
\end{strip}

\noindent
observables. For example, the projection of ${T}_{\mu \nu }$ onto a timelike direction is

\begin{equation}
{T}_{\mu \nu }{{u}^{\mu }}{{u}^{\nu }}=\varepsilon,
\end{equation}
and the projection of ${{T}_{\mu \nu }}$ onto $3D$ plane is
\begin{equation}
{{T}_{\mu \nu }}\,{{b}^{\mu \lambda }}=-p\,b_{\nu }^{\lambda }.
\end{equation}

In our case, taking into account relation (4), the set of field equations can be reduced to the equation 

\begin{equation}
{\sigma }''+((2+3\beta )/S)\cdot {\sigma }'+((1+3\beta )/2)\cdot {{({\sigma }')}^{2}}=0,
\end{equation}
where the prime denotes a derivative with respect to $S.$ 
If we substitute

\begin{equation}
\sigma =\frac{2}{(1+3\ \beta )}\cdot \ln \,\varphi, 
\end{equation}
then (7) becomes 

\begin{equation}
{{\varphi }'}'\,+\frac{(2+3\,\beta )}{S}\cdot {\varphi }'\,=0,
\end{equation}
which is the Laplace equation of multidimensional Euclidean spherical space if we take $\beta$ as discrete quantity (rational number)
\begin{equation}
2+3\beta =k-1,\quad \quad \beta =\frac{k-3}{3}
\end{equation}
with $k$ as an integer. 

In this case $k$ is a dimension of a multidimensional Euclidean space, $k=0,1,2,..,6$. Furthermore from the energy dominance principle $|\beta | \leq 1$ we can construct Tabble 1 for comparison with corresponding states of matter.

So we now rewrite (9) as
\begin{equation}
{{{{\varphi }'}'}_{\ (k)}}+\frac{(k-1)}{S}\cdot {\varphi }'{{\,}_{(k)}}=0
\end{equation}
or in a compact form
\begin{equation}
\Delta {}_{k}\ {{\varphi }_{(k)}}=0,
\end{equation}
where ${{\varphi }_{(k)}}$ are functions of $S$ for different values of $k$ or discrete values of the parameter $\beta;\; $ $\Delta {}_{k} $ is the Laplace operator in multidimensional Euclidean spherical 
$k$-dimensional space (a radial part of the multidimensional Laplace operator). 

The values of $k$ are restricted by the energy dominance principle, $\,|\beta | \leq 1$. The index in parentheses points at the dimension of space in which we consider our quantities. For example, the explicit form of the function ${{\varphi }_{(k)}}$ depends on $k$. 

Thus we can replace the cosmological problem in pseudo-Riemannian 4D space-time 
with a study in auxiliary $k$-dimensional Euclidean space.

\section{\large{\tt FORMULATION OF A VARIATIONAL \\
PROBLEM IN MULTIDIMENSIONAL \\
EUCLIDEAN SPACE}}

We have now the set of Euclidean $k$-dimensional spaces ${{M}_{k}}$ where the variable $S$ plays the role of the radial variable. This variable can be written using Cartesian coordinates ${{\xi }^{A}}$ in this space

\begin{equation}
{{S}^{2}}={{\eta }_{AB}}\,\xi {{\,}^{A}}\,\xi {{\,}^{B}},
\end{equation}
where ${{\eta }_{AB}}=
diag(1,\,1,\,1,\,..,1)\,;\ \ A,\,B,\,..,\,D\ = \,\ 1,\,2,\,  \newline 
..,\,k$ and in this space a time-like coordinate is absent. 

Now we introduce the differential 1-form
\begin{multline}
{{\bf E}_{(k)}}= \\
=-({{\nabla }_{A}}\,{{y}^{(k)}}\,)\,{\bf d}\xi {{\,}^{A}}\equiv 
\\
-\,{{y}^{(k)}}{{,}_{A\,}}{\bf d}\xi {{\,}^{A}}\ \equiv \ E_{A}^{(k)}\,{\bf d}{{\xi }^{A}},
\end{multline}
where in Euclidean space we can write ${{\nabla }_{A}}=\partial /\partial {{\xi }^{A}} \; ; \newline
\ {{y}^{(k)}}={{y}^{(k)}}(S).$ This differential 1-form is closed, i.e., ${\bf d}{{E}_{(k)}}\ =\ 0 $ 
(${\bf d}\, $ is an external differential). We set 
$$
|{{\bf E}_{(k)}}{{|}^{2}}={{E}_{A}}{{E}^{A}} = 
{{\eta }^{AB}}\,{{\nabla }_{A}}{{y}^{(k)}}{{\nabla }_{B}}{{y}^{(k)}}
$$ 
and write the ''energy'' functional in the form

\begin{multline}
{{E}^{(k)}}=
\\
=\alpha _{0}^{(k)}\int\limits_{{{V}_{(k)}}}{|{{E}_{(k)}}{{|}^{2}}d{{V}_{(k)}} = 
 \alpha _{0}^{(k)}\,\int\limits_{{{V}_{(k)}}}{(\,{{E}_{A}}{{E}^{A}}\,)d{{V}_{(k)}}}} \\
=\alpha _{0}^{(k)}\,\int\limits_{{{V}_{(k)}}}{{{\eta }^{AB}}\,
{{\nabla }_{A}}{{y}^{(k)}}{{\nabla }_{B}}{{y}^{(k)}}\,\,d{{V}_{(k)}}},
\end{multline}
where $\{\alpha _{0}^{(k)}\}$ are constant coefficients which simulate the corresponding 
``permittivities'', ${{V}_{(k)}}$ is $k$-dimensi- \newline 
onal integration volume in ${{M}_{(k)}}$ for 
$k=3$ $\{{{E}_{A}}\}=({{E}_{1}},{{E}_{2}},{{E}_{3}})=\vec{E}$ is an analog of the electric field intensity. 

The requirement of a minimum of such a functional (${{\delta }_{v}}{{E}^{(k)}}=0 $ ) leads to the Laplace equation
\begin{equation}
({{\eta }_{AB}}\,{{\nabla }_{A}}{{\nabla }_{B}}){{y}^{(k)}}={{\Delta }_{(k)}}{{y}^{(k)}}=0,
\end{equation}
or
\begin{equation}
{{\nabla }_{A}}\,{{E}^{A}}=0.
\end{equation}

This equation corresponds to Maxwell's equation $div\,\vec{E}=0$ in the absence of charges ($\vec{E}\,$ is the electrical field intensity) for $k=3,\,$ and Eq.(16) is, on the other hand, identical to 
Eqs.(11) and (12) , i.e., the differential operator ${{\Delta }_{(k)}}$ is the radial part of the Laplace equation in $k$-dimensional Euclidean space.

Thus, the cosmological problem with conformally flat metric (1) and with the equation of state (4) can be replaced with a variational problem similar to minimization of the energy of an electrostatic potentials distribution in multidimensional Euclidean space. In this case, functions 
${{y}^{(k)}}(S)$ play the role of electrostatic potentials, and ${{E}_{A}}$ are an analog of the electrostatic field intensity in ${{M}_{(k)}}.$ 

\section{\large{\tt INTRODUCTION TO THE CONCEPT \\ 
OF MULTIDIMENSIONAL CAPACITORS}}

In the future we will keep in mind the presence of two ``conductors'' which act as capacitor plates and restrict the spatial part ${{M}_{(k)}}$. These ``plates'' are the hypersurfaces of dimensionality smaller than that of ${{M}_{(k)}}.$ We denote these hypersurfaces as ${{\Sigma }_{(k)}}$ with dimensionality ($k-1$ ).

Now we will demand boundary conditions as the $3D$ analog 
\begin{equation}
{{y}^{(k)}}_{|\Sigma _{(k)}^{(1)}}-{{y}^{(k)}}_{|\Sigma _{(k)}^{(2)}\ }={{U}^{(k)}},
\end{equation}
where ${{U}^{(k)}}$ are analogs of potential difference (voltages); 
${{U}^{(k)}}$ are taken on different conductors. The ``potentials'' on the hypersurfaces $\Sigma _{(k)}^{(1)}$ and $\Sigma _{(k)}^{(2)}$ are constants.

Further we will use the identity

\begin{multline}
{{\eta }^{AB}}{{\nabla }_{A}}{{y}^{(k)}}{{\nabla }_{B}}{{y}^{(k)}}=
\\
-{{y}^{(k)}}({{\eta }^{AB}}{{\nabla }_{A}}{{\nabla }_{B}}){{y}^{(k)}}+{{\eta }^{AB}}{{\nabla }_{A}}({{y}^{(k)}}{{\nabla }_{B}}{{y}^{(k)}})
\end{multline}
and a generalization of the Gauss theorem (${{\bf F}^{A}}$ is some vector)
\begin{equation}
\int\limits_{{{V}_{(k)}}}{({{\nabla }_{A}}{{F}^{A}})\,d{{V}_{(k)}}=}\oint\limits_{{{\Sigma }_{(k)}}}{({{F}^{A}}{{n}_{A}})\,d{{\Sigma }_{(k)}}},
\end{equation}
where ${{n}_{A}}$ is a normal unit vector to the hypersurface, ${{n}_{A}}{{n}^{A}}=1.$

Then we can transform the expression (15) to 
\begin{multline}
{{\bf E}^{(k)}}/\alpha _{0}^{(k)} = \int\limits_{{{V}_{(k)}}}{{y}^{(k)}}(\,{{\nabla }_{A}}{{E}^{A}}\,)d{{V}_{(k)}}+
\\
\oint\limits_{{{\Sigma }_{(k)}}}{{{y}^{(k)}}({{\eta }^{AB}}{{n}_{B}}{{\nabla }_{A}}{{y}^{(k)}}})d{{\Sigma }_{(k)}}.
\end{multline}

The volume integral disappears taking due to (17). The second integral is taken over the 
``plates'' (hypersurfaces), i.e. ${{\Sigma }_{(k)}}=
\Sigma _{(k)}^{(1)}+\Sigma _{(k)}^{(2)}$. So this integral over the hypersurface taking into account the boundary condition (18), is  transformed into 

\begin{multline}
{{\bf E}^{(k)}}/\alpha _{0}^{(k)}={{y}^{(k)}}_{|\Sigma _{(k)}^{(1)}}\cdot \oint{({{n}^{A}}{{\nabla }_{A}}{{y}^{(k)}})\,d\Sigma _{(k)}^{(1)}-}
\\
{{y}^{(k)}}_{|\Sigma _{(k)}^{(2}}\cdot \oint{({{n}^{A}}{{\nabla }_{A}}{{y}^{(k)}})\,d\Sigma _{(k)}^{(2)}}.
\end{multline}

In electrostatics with $k=3,$ the electric field intensity vector $\vec{\bf E}$ is taken on a closed surface around an electrical charge. This surface is equipotential, and there is the true equality $(\vec{\bf E}\cdot \vec{n})\cdot {{\Sigma }_{(3)}}=-\,4\pi \cdot {{Q}_{(3)}}$, where ${{\Sigma }_{(3)}}=4\pi \cdot {{S}^{2}}.$ 

In the general case ($ k>2 $) we can write

\begin{equation}
{{n}^{A}}{{\nabla }_{A}}{{y}^{(k)}}={{\alpha }_{(k)}}{{Q}_{(k)}}(k-2),
\end{equation}
where ${{Q}_{(k)}}$ are analogs of charges, while the coefficients ${{\alpha }_{(k)}}$ are here
\begin{equation}
{{\alpha }_{(k)}}=
\left \{\begin{aligned}
  & k{{\pi }^{b}}/b! ,\quad \quad \quad \quad \quad k=2b, \\ 
 & \frac{{{\pi }^{b}} b!{{2}^{k}}}{(k-1)!} , \quad\quad \quad\quad \; \;  k=2b+1,  
\end{aligned} \right \}
\end{equation}
where $b=1,\,2,....$ Then the expression (22) is rewritten as 
\begin{equation}
{{\bf E}^{(k)}}=\alpha _{0}^{(k)}\cdot {{\alpha }_{(k)}}{{U}_{(k)}}{{Q}_{(k)}}(k-2).
\end{equation}

By analogy with the 3D space, we define the capacity of two conductors as the coefficient linking the electric charge and the electric potential difference:
\begin{equation}
{{Q}_{(k)}}={{C}_{(k)}}\cdot {{U}_{(k)}},
\end{equation}
where ${{C}_{(k)}}$ are analogs of the electrical capacities in multidimensional space.

Finally we get the equality
\begin{equation}
{{\bf E}^{(k)}}=(k-2)\cdot \alpha _{0}^{(k)}\cdot {{\alpha }_{(k)}}U_{(k)}^{2}\,{{C}_{(k)}}
\end{equation}

Therefore the expression (27) establishes a minimum value of the system capacity for given boundary conditions. We must here say that the capacity of the capacitor depends on both the shape and the size of the conductors.


\section{\large{\tt MULTIDIMENSIONAL CAPACITY, \\ 
HARMONIC FUNCTIONS \\
 AND COSMOLOGICAL MODELS}}

Now we can find ${{C}_{(k)}}$ by using the expression (26) 
and ${{\bf E}^{(k)}}$ from (27) as 
\begin{equation}
{{C}_{(k)}}=(k-2)\cdot \alpha _{0}^{(k)}{{\alpha }_{(k)}}\cdot Q_{(k)}^{2}/{{E}^{(k)}}.
\end{equation}

If the functions ${{y}^{(k)}}$ are solutions of the Laplace equation (16), we can compute the energy functional ${{E}^{(k)}}$ for closed shells (``balls'') ${{\Sigma }_{(k)}}$ very easily. The operator 
${{\Delta }_{(k)}}$ is only a radial part of total the Laplace operator. The solutions 
${{y}^{(k)}}$ of such an equation are well known as harmonic functions:
\begin{equation}
{{y}_{(k)}}={{A}_{(k)}}/{{S}^{k-2}}\,;\quad \quad \quad k\ge 3,
\end{equation}
but the derivatives of these functions are not solutions of the Laplace equations. Here ${{A}_{(k)}}$  are analogs of electric charges on the plates. 

We will first substitute the derivatives of the function ${{y}^{(k)}}$ into 
(15). Then we will integrate over space between the shells for $k>2.$ After using 
(18) and that $d{{V}_{(k)}}={{\alpha }_{(k)}}{{S}^{k-1}}dS $ we get the following expression for a multidimensional ``ball'' capacitor: 
\begin{equation}
{{E}^{(k)}}=\alpha _{0}^{(k)}\cdot {{\alpha }_{(k)}}(k-1){{A}_{(k)}}{{U}_{(k)}}.
\end{equation}

If we now ''ground'' one of the capacitor plates, the total electric charge of the capacitor will be equal to zero, and ${U}_{(k)}$ will be written as the potential inside the ''ball'': 
\begin{equation}
{{U}_{(k)}}={{\Phi }_{(k)}}={{A}_{(k)}}(1-/S_{1}^{k-2}-1/S_{2}^{k-2}).
\end{equation}

Substituting the expression (31) into (30) and choosing 
$\ {{Q}_{(k)}}\equiv {{A}_{(k)}},\, $ we have from (eq28)
\begin{multline}
{{C}_{(k)}}=\alpha _{0}^{(k)}{{\alpha }_{(k)}}{{Q}_{(k)}}/{{\Phi }_{(k)}}=
\\
\alpha _{0}^{(k)}{{\alpha }_{(k)}}{{(1-/S_{1}^{k-2}-1/S_{2}^{k-2})}^{-1}}
\end{multline}
or 
\begin{equation}
{{C}_{(k)}}{{\Phi }_{(k)}}=\alpha _{0}^{(k)}{{\alpha }_{(k)}}{{Q}_{(k)}}.
\end{equation}

Now choosing the ``permittivity'' so that with 
$C_{(k)} \linebreak
\to\alpha _0^{(k)}{{\alpha }_{(k)}}S_{1}^{k-1}\equiv {{q}_{(k)}},$ i.e. 
$\alpha _{0}^{(k)}= 
{{q}_{(k)}}/({{\alpha }_{(k)}}S_{1}^{k-2})$ we get the expression 
\begin{equation}
{{C}_{(k)}}={{q}_{(k)}}{{(1-{{({{S}_{1}}/{{S}_{2}})}^{k-2}})}^{-1}}.
\end{equation}

Further we fix ${{S}_{1}}$ and admit changes of the radius ${{S}_{2}}\equiv S.$ In this case we have 
${{C}_{(k)}}={{C}_{(k)}}(S),$ and the expression (34) may be rewritten as 
\begin{equation}
{{C}_{(k)}}(S)\cdot {{\varphi }_{(k)}}(S)={{q}_{(k)}},
\end{equation}
where we require that the functions ${{\varphi }_{(k)}}(S)$ are identical to the functions which are solutions of Eqs. (11) and (12)). We immediately find for $3 \leq k \leq 6$ 
\begin{equation}
e^{2\sigma} = {{({{\varphi }_{(k)}})}^{4/(k-2)}}=
{{({{q}_{(k)}}/{{C}_{(k)}})}^{4/(k-2)}},
\end{equation}
i.e., the existence of a conformally flat metric for open cosmological models with special equations 
of state (see (4)) can be connected with a computation of the capacity of multidimensional 
ball capacitor and then be used in the expression (36).

If in Eq. (32) the value of ${{S}_{2}}$ tends to infinity, we get the capacity of an isolated body
\begin{equation}
{{C}_{(k)}}=({{q}_{(k)}}/{{\tilde{q}}_{(k)}}){{S}^{k-2}},
\end{equation}
where now $\alpha _{0}^{(k)}{{\alpha }_{(k)}}={{q}_{(k)}}/{{\tilde{q}}_{(k)},} and {{S}_{1}}\equiv S .$ The expression (36) is transformed in this case to 

\begin{equation}
e^{2\sigma}={{(({{q}_{(k)}}/{{\tilde{q}}_{(k)}})/{{S}^{(k-2)}})}^{4/(k-2)}}.
\end{equation}

Further on we need to consider separately the variants of the conformal factor for degenerate cases with $k=0,\ 1,\ 2$.


\section{\large{\tt SOME EXCEPTIONS \\
FROM THE GENERAL APPROACH }}

The above was studied for metrics with asymptotically flat at infinity ($S\to \infty $). For models with the equations of state $\beta =-1,\,-2/3,\,-1/3$ 
(see Table 1), which correspond to $k=0,\ 1,\ 2$ we must return to the original equation 
(16). 

1. The physical state with $\beta =-1$ corresponds to $k=0$ and we cannot integrate the expression 
(15), but the solution of Eqs. (11) and (12) in this case describes the 
de Sitter model of the physical vacuum in the coordinates chosen here 
(see [1], [3], [10] and Table 1). 

2. In the one-dimensional auxiliary Euclidean space ($k=1$, $\beta =-2/3$) we have 
${{y}_{(k)}}={{B}_{(k)}}+{{A}_{(k)}}S $ and ${{E}^{(1)}}=\alpha _{0}^{(1)}A_{(1)}^{2}({{S}_{2}}-
{{S}_{1}})$. Then the expression for ${{C}_{(1)}}\propto 1/({{S}_{1}}-{{S}_{2}})=1/d$ can be interpreted as the capacity of a plane capacitor with a distance between plates 
$d=({{S}_{2}}-{{S}_{1}})$. If we take this distance as a variable quantity 
(${{S}_{2}}\equiv S>{{S}_{1}}$), then ${{C}_{1}}={{C}_{1}}(S)$ and 
$e^{2\sigma} = {{({{q}_{(1)}}/{{C}_{(1)}})}^{-4}}\propto 1/{{d}^{4}}$ (see [1], [3] and Table 1). The Euclidean asymptotic form of this metric will be naturally broken. 

3. When $k=2,$ i.e., $\beta =-1/3$, we must compare Eqs. (7) and 
(17) by putting $\sigma \equiv {{y}_{(2)}}$ with ${{y}_{(2)}}=
{{B}_{(2)}}\,\ln ({{A}_{(2)}}\cdot S)$. The computations in the 2D case lead to 
${{\bf E}^{(2)}}=2\pi \cdot \alpha _{0}^{(2)}B_{(2)}^{2}\ln ({{S}_{2}}/{{S}_{1}}).$ Finally, we have the capacity ${{C}_{2}}=({{Q}_{(2)}}/\lambda ){{(\ln ({{S}_{2}}/{{S}_{1}}))}^{-1}}$ for a cylindrical capacitor with $\lambda = 2\pi \cdot \alpha _{0}^{(2)}B_{(2)}^{2}/{{Q}_{(2)}}$. In this case we can have the different cosmological models depending on the choice of the value  of $\lambda $ and of the variability of the capacitor plate radius. 

For example, if ${{S}_{2}}\equiv S $ and $\lambda =2 ,$ we find $e^{2\sigma} =
{{(S/{{S}_{1}})}^{4}}.$ This means that, at fixed spatial variable, such model is the Dirac cosmological model [10]. If we assume that ${{S}_{2}}>>{{S}_{1}}=S $ and $\lambda =1$, then 
$e^{2\sigma} = {{({{S}_{2}}/{{S}_{1}})}^{2}}$. The metric with such a conformal factor describes the Milne universe [10] in the interval $0<S<{{S}_{2}}.$ 


\section{\large{\tt ASYMPTOTICALLY FLAT METRICS }}

We first start with the case of the $k=3$ ($\beta =0,$ see Table 1). From 
(34) we have a spherical capacitor in the ancillary Euclidean space. Its capacity is connected with conformal factor of the metric (1), as we can see from the expression (36). 

The corresponding conformal factor of the open Friedman Universe [7] is:
\begin{equation}
e^{2\sigma}={{(1-{{A}_{F}}/S)}^{4}},
\end{equation}
where ${{S}_{1}}\equiv {{A}_{F}}$ is the Friedman constant which can be found from the Friedman asymptotic form. 

If we will take $k=4$ ($\beta =1/3,$ see Table 1), then it is easily to find 
\begin{equation}
e^{2\sigma}={{(1-B/{{S}^{2}})}^{2}},
\end{equation}
where ($ S_{1}^{2}=B=const $). This metric describes an open cosmological model filled with 
equilibrium radiation with the equation of state $3p=\varepsilon $. 

In the cases $k=5$ ($\beta =2/3$) and $k=6$ ($\beta =1$) we get the conformal factors with the constants $C$ and $D$ (compare with Table 1) 
\begin{equation}
e^{2\sigma} = {{(1-C/{{S}^{3}})}^{4/3}}
\end{equation}
and
\begin{equation}
e^{2\sigma} \quad =\quad (1-D/{{S}^{4}}).
\end{equation}

\section{\large{\tt CONCLUSION}}

The above demonstration of the analogy between open cosmological models in 4D space-time and the 
``ball'' electric capacitors in multidimensional Euclidean spaces of dimensions $k=1,\ 2,\ 3,..,\ 6$ shows the opportunity of replace the simulation problem in cosmology with the equivalent electrostatic problem for the finding the corresponding capacities under given boundary conditions. 

In addition, the expressions of the conformal factors for $k=4,\ 5,\ 6$ are generalizations of the Friedman solution (39). 

In fact, if we assume the validity the Einstein equations in space-times of other dimensions with 
$m = N+1 > 4$ as is done in [2,3], then a generalization of the equation 
(11) will be
\begin{equation}
{\Phi }^{\prime \prime}+((\nu -1)/S)\ {\Phi }^{\prime}=0,
\end{equation}
where $\,\nu =N(1+\beta )\,;\quad N$ is the dimension of a spatial hypersurface orthogonal to the 
timelike congruence, which is a generalisation of 3D space; here 
 $\sigma =2(\ln \,\Phi )/(\nu -2).$ If $\nu $ runs through integer values 
$\nu \equiv n = N(1+\beta ),$ we immediately have $\beta =0$ for those Euclidean spaces which have dimensions equal to $n=N\,$ (see Table 2).

Table 2 shows the range of the quantity $\beta$ in space-times of dimensions $m$ higher than $4.$ 
Each value $\beta$ corresponds to a harmonic function which is the solution of the radial Laplace equation (43). And each harmonic function corresponds to the conformal factor of the cosmological metric.

The full picture is clearly reflected in Table 2, which is filled by values of $\beta,$ corresponding to various multidimensional space-times. It is easy to see that only three
 ``states of matter'' are present in all dimensions: ``physical vacuum'', ``incoherent dust'' and ``superstiff state''. These terms are in quotation marks because physical interpretations of specific meaning of these terms in spaces of different dimensions requires a separate consideration. In this regard, there is an interesting fact that in the 4D world (the third column) all possible values of the parameter $\beta$  have a clear physical interpretation and include all essential states of matter used to find cosmological solutions.

It is also necessary to point out that harmonic functions of the same order (in other words, with the same conformal factor) correspond to different values of $\beta $ in spaces of different dimensions. 
In this case, can speak of a ``unification'' of the interaction types because for any $\beta $ 
(except physical vacuum) for given space-times there is such a space-time with another dimension in which the corresponding conformal factor describes a model filled with matter of the type of dust 
with $\beta =0 .$ 

\newpage

\begin{strip}

{\bf Table 2.} The quantity $\beta = p/{\varepsilon}$ in space-time of dimensions higher than 4
\vspace{2mm}
\begin{center}
\begin{tabular} {|l|c|c|c|c|c|c|c|c|c|c|c|c|r|}  \hline
$N/n$ & 1 & 2 & 3 & 4 & 5 & 6 & 7 & 8 & 9 & 10 & 11 & 12 & ...        \\ \hline
 0   &-1  &-1 &-1 &-1 &-1 &-1 &-1 &-1 & -1 & -1 & -1 & -1 & ...        \\ 
 1   & {\bf0}  &-1/2 &-2/3 &-3/4 &-4/5 &-5/6 &-6/7 &-7/8 & -8/9 & -9/10 & -10/11 & -11/12& .\\ 
 2   &+1 & {\bf0} &-1/3 &-2/4 &-3/5 &-4/6 &-5/7& -6/8 & -7/9 & -8/10 & -9/11 & -10/12& ... \\ 
 3   &   &  +1/2 &  {\bf0}   &-1/4  &-2/5 &-3/6 & -4/7& -5/8&-6/9&-7/10&-8/11&-9/12&... \\ 
 4   &   & +1  &+1/3& {\bf0} &-1/5& 2/6&-3/7&-4/8&-5/9&-6/10&-7/11&-8/12&...   \\ 
 5   &   &     &+2/3 & +1/4&{\bf0}&-1/6&-2/7&-3/8&-4/9&-5/10&-6/11&-7/12&... \\ 
 6   &   &   & +1 & +2/4&+1/5&{\bf0}&-1/7&-2/8&-3/9&-4/10&-5/11&-6/12&...    \\ 
 7   &   &   &  & +3/4&+2/5&+1/6&{\bf0}&-1/8&-2/9&-3/10&-4/11&-5/12&...      \\ 
 8   &   &   &  & +1&+3/5&+2/6&+1/7&{\bf0}&-1/9&-2/10&-3/11&-4/12&...    \\ 
 9   &   &   &  &  &+4/5&+3/6&+2/7&+1/8&{\bf0}&-1/10&-2/11&-3/12&...   \\ 
10   &   &   &  &  &+1 &+4/6&+3/7&+2/8&+1/9&{\bf0}&-1/11&-2/12&....     \\ 
11   &   &   &  &  &   &+5/6&+4/7&+3/8&+2/9&+1/10&{\bf0}&-1/12&...   \\ 
12   &   &   &    &   & &+1 &+5/7&+4/8& +3/9&+2/10&+1/11&{\bf0}&...    \\ 
13   &   &   &  & &  &  & +6/7 & +5/8&+4/9& +3/10&+2/11&+1/12& ...   \\ 
14   &   &   &  & &  &  &...&...&...&...&...&...&...                   \\ \hline
\end{tabular}
\end{center}

\vspace{3mm}

\end{strip}

Moreover Table 2 reflects a more general consideration of the analogy between multidimensional condensers and models of the open Universe via sets of values $\beta$ and the order of harmonic functions.

\vspace{.2cm}
\small

\begin{center}
{\large{\tt REFERENCES}}
\end{center}

1. A.M.Baranov and E.V.Saveljev,``Conformal-flat models and equations of state. 1.The four-dimensional space-time'', Dep.VINITI USSR, 05.07.88, no 5914-B88 (1988) (in Russian).

2. A.M.Baranov and E.V.Saveljev, ``Conformal-flat models and equations of state. 2.The multidimensional space-times'', Dep.VINITI USSR, 18.04.90, no. 2096-B90 (1990) (in Russian).

3. A.M.Baranov and E.V.Saveljev, ``Mutidimensional con-\newline 
formal-flat space-times and a linear equation of state'', {\it Physical Interpretation of Relativity theory}, Proc. Int. Meeting, Bauman Moscow State Technical University, Moscow, 29 June 2015 (BMSTU, Moscow, 2015), p.81-100 (2015), doi: 10.18698/2309-7604-2015-1-81-100.

4. A.M.Baranov and E.V.Saveljev, ``Spherically symmet- \newline 
ric lightlike radiation and conformally flat space-times'', Izv. vuz.(Fizika) 7, 32-35 (1984).

5. A.M.Baranov and E.V.Saveljev, ``Spherically symmetric lightlike radiation and conformally flat space-times'', Russ. Phys. J., {\bf 27}, 569-572 (1984).

6. A.M.Baranov and E.V.Saveljev, ``Exact solutions of the conformal-flat Universe. I. The evolution of model as the problem about a particle movement in a force field'', Space, Time and Fundamental Interactions (STFI) 1, 37-46 (2014).

7. V.A.Fock, {\it The Theory of Space, Time and Gravitation},  (New York: Pergamon, U.S.A.,2nd edition, 1964 ).

8. Yu.S.Vladimirov {\it Reference Frames in the Gravitation Theory} (Moscow: Energoizdat, 1982), (in Russian).

9.N.V.Mitskievich  {\it Relativistic Physics in Arbitrary Reference Frames} (New York: Nova Science Publishers Inc., 2006).

10. L.Infeld L. ``A New Approach to Kinematic Cosmology'', Phys.Rev., {\bf 68},
250-272 (1945).

\end{document}